\documentclass[conference]{IEEEtran}
\IEEEoverridecommandlockouts
\usepackage{cite}
\usepackage{amsmath,amssymb,amsfonts}
\usepackage{graphicx}
\usepackage{paralist}
\usepackage{textcomp}
\usepackage{bm,amsthm}
\usepackage{xcolor}
\usepackage[ruled,vlined]{algorithm2e}
\usepackage{paralist}

\def\BibTeX{{\rm B\kern-.05em{\sc i\kern-.025em b}\kern-.08em
    T\kern-.1667em\lower.7ex\hbox{E}\kern-.125emX}}
\theoremstyle{definition}
\newtheorem{exmp}{Example}[]
\SetVlineSkip{0pt}

\begin{document}

\title{Automated Malware Design for Cyber Physical Systems
\thanks{}
}

\author{\IEEEauthorblockN{Ashraf Tantawy}
\IEEEauthorblockA{\textit{Department of Electrical and Computer Engineering} \\
\textit{Virginia Commonwealth University}\\
Richmond, VA, USA \\
ORCID: 0000-0003-0958-1633}
}

\IEEEoverridecommandlockouts

\maketitle

\IEEEpubidadjcol

\begin{abstract}
The design of attacks for cyber physical systems is critical to assess CPS resilience at design time and run-time, and to generate rich datasets from testbeds for research. Attacks against cyber physical systems distinguish themselves from IT attacks in that the main objective is to harm the physical system. Therefore, both cyber and physical system knowledge are needed to design such attacks. The current practice to generate attacks either focuses on the cyber part of the system using IT cyber security existing body of knowledge, or uses heuristics to inject attacks that could potentially harm the physical process. In this paper, we present a systematic approach to automatically generate integrity attacks from the CPS safety and control specifications, without knowledge of the physical system or its dynamics. The generated attacks violate the system operational and safety requirements, hence present a genuine test for system resilience. We present an algorithm to automate the malware payload development. Several examples are given throughout the paper to illustrate the proposed approach.
\end{abstract}

\begin{IEEEkeywords}
Cyber Physical System, CPS, Attack, Malware, Safety, Security, Control System, SCADA, Formal Specification, Safety Property.
\end{IEEEkeywords}

\section{Introduction and Related Work}
A Cyber-Physical System (CPS) is an integration of computation, networking, and physical processes to monitor, control, and safeguard the physical system \cite{Rajkumar2010}. Most cyber physical systems are mission-critical, such as process control plants, energy grid, autonomous vehicles, and pacemakers, just to name a few \cite{Lee2017}. These systems progressively replace closed-source hardware and software components with open source embedded system designs, operating systems, and standard communication protocols, mainly to simplify integration efforts and to reduce total cost of ownership. This open-source paradigm has exposed, and continues to expose, these mission critical systems to a myriad of security threats \cite{Langner2011a}. When such threats materialize into a successful attack, the outcome could be catastrophic in terms of human casualty, environmental damage, and financial loss.

To protect against CPS cyber threats, system designers need to be at least one step ahead of attackers. This requires a rigorous security risk assessment process at different stages of the design process \cite{Tantawy2020}. A critical component of security risk assessment is penetration testing, where a set of attacks are launched to assess system vulnerabilities and consequence severity. For penetration tests to be effective, attacks should be comprehensive and non-generic, i.e., closely-related to the system under study. The current penetration testing practice relies on the pen tester expertise as well as IT security type attacks. Little work has been done on how to design attacks for a given cyber physical system.

In this work, we make an initial step towards automating malware design for cyber physical systems. The idea is to extract system knowledge embedded in control and safety software programs used to safeguard the system. By simple reversal of the system operating logic, we guarantee the generation of a relevant system attack with a consequence varying from operation disruption to a system hazard. Little pre-requisite knowledge is required to design such attacks. Therefore, they could be utilized by penetration testers to study the impact of viable attack scenarios against the CPS under consideration.

Research on attack design for cyber physical systems is a relatively new field. Most of the work so far has focused on modeling the attacker behavior and objectives, collectively referred to as \emph{attacker profile} \cite{Adepu2016a, Rocchetto2016a, Orojloo2016, Deloglos2020}. This work attempts to couple the attacker objectives to the cyber physical system in an abstract way that is applicable to any given CPS. Translation of abstract attacker models into actual attacks that are relevant to the given CPS is not yet explored. Another line of research is the development of specialized attacks for CPS protocols, such as Modbus and DNP3 protocols \cite{Chen2015,Singh2014, East2009, Jin2011, Darwish2015, Huitsing2008, Bashendy2020}. This work is important but still lacks a systematic way to include physical system knowledge into the attacks. A related emerging research is the integration of security and safety specifications in the design process rather than treating security as a post-design issue \cite{Kriaa2015b,Lyu2019}. The automated generation of cyber attacks from the model-based engineering design process is still unexplored. 

The work in this paper distinguishes itself in that it utilizes the physical system knowledge in a systematic way to design integrity attacks customized to the given cyber physical system, an activity that is currently done in an ad-hoc manner. We utilize system specifications expressed in formal logic, which is a well-established domain \cite{Lee2017}. We summarize the key contributions of the paper as follows: \begin{inparaenum}[(1)] \item A formal method to design a malware for a given cyber physical system, \item Automated generation of malware payload given a CPS safety program and optionally a control program, \item Inference of system dynamic relationships from the control algorithms without knowledge about system model, and \item A systematic process for attack injection. \end{inparaenum}

\begin{figure}[tb!]
    \centering
    \includegraphics[scale=0.65]{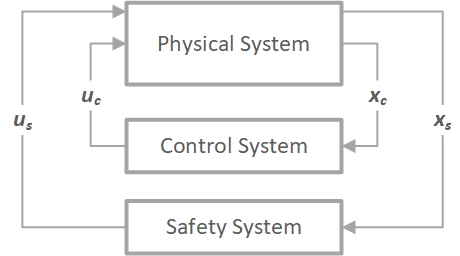}
    \caption{Cyber Physical System Architecture}
    \label{fig:System-Arch}
\end{figure}

Figure \ref{fig:System-Arch} shows the classical cyber physical system architecture that is utilized in the paper. The control system typically controls the normal operation of the system and keeps the state variables within the pre-designed operating envelope. The safety system is an independent embedded system, with its own sensors and actuators. The safety system is normally dormant and intervenes only when the the control system is incapable of keeping the state variables inside the operating envelope. The intervention is mostly in the form of a system emergency shutdown to prevent any existing hazard from developing into a catastrophic event.

The paper is organized as follows: Section \ref{sec:SafetyReq} is a brief introduction to safety system requirements and their implementation in software. Section \ref{sec:AttackDesign} describes the attack design methodology, including how to infer plant dynamics from the control software. Section \ref{sec:AttackInjection} briefly explains the attack injection process and how the presented work fits into the big picture of penetration testing. The work is concluded in Section \ref{sec:Conclusion}. Instead of presenting a case study at the end of the paper, we utilize running examples throughout the paper to avoid abstract discussion and to ensure effortless reading.

\section{Safety Requirements Specification} \label{sec:SafetyReq}
A safety property for a cyber physical system is generally formulated as a temporal logic formula \cite{Peled1995}. One of the simplest and most common forms of temporal logic formulae is the invariant; a property that holds true at all times during system operation \cite{Lee2017}. The invariant is often expressed in natural language and translated into a logical expression for software or hardware implementation. An example invariant for a traffic control system is "\emph{Pedestrian crossing light must be off when the traffic light is green}". Another example invariant for a process control plant could be "\emph{The pump should be stopped when the tank level is low}". System properties for a CPS could be defined for both the physical and cyber system. An example invariant for a multi-threaded C language code running on a controller is "\emph{The program should never get into a deadlock}". We focus in this work on safety properties of the physical system, since this is what distinguishes CPS security from IT security domain.

In practice, an invariant is ensured by implementing a logic formula that relates a subset of system state variables to a system output (actuator). The state variables could represent sensor measurements, human input, or environmental state. The invariant property could be expressed by the implication $(P(\bm{x}) \Longrightarrow u)$, where $P$ is a propositional function, $\bm{x}$ is a subset of the state variables, and $u$ is an actuator output. Since this is an invariant, at any time of system operation and at any state, whenever $P(\bm{x})$ is true, $u$ must be true. The invariant could be implemented in software using an IF THEN construct: $\texttt{IF } P(\bm{x}) \texttt{ THEN } u=1$.

\begin{figure}[tb!]
    \centering
    \includegraphics[scale=0.6]{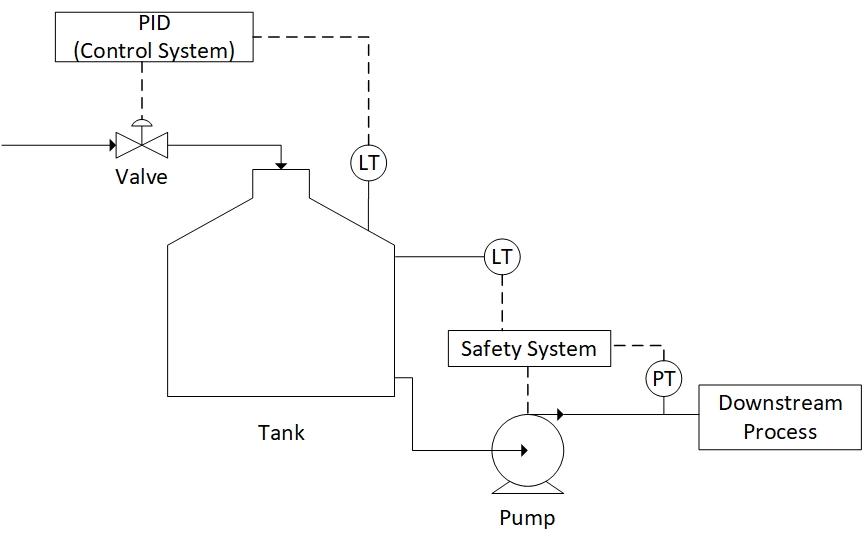}
    \caption{A tank-pump system with pump safety defined by tank level and discharge pressure.}
    \label{fig:Tank-Pump}
\end{figure}

\begin{exmp}\label{ex:pump-control}
Consider the simple tank-pump system in Figure \ref{fig:Tank-Pump}. The pump safety property: "\emph{The pump should stop if the tank level is below 10\% or the pump discharge pressure is above 5 bars}". Let $x_1$ denote the tank level, $x_2$ denote the pump discharge pressure, and $u$ denote the pump output, where $u=1$ when the pump is running. The invariant property is then expressed as:
\begin{align}
    \underbrace{(x_1 < 10\%) \lor (x_2 > 5)}_{P(\bm{x})} \Longrightarrow \lnot u
\end{align}
and the software implementation will be:
\begin{align}
    & \texttt{IF } (x_1 < 10\%) \lor (x_2 > 5) \texttt{ THEN } u=0  \tag*{$\blacksquare$}
\end{align}

\end{exmp}

\section{Attack Design} \label{sec:AttackDesign}
The invariant expression $(P(\bm{x}) \Longrightarrow u)$ defines the logic for $u=1$, but assumes the logic for $u=0$ is defined elsewhere by another logic. For attack design purposes, we utilize the bi-conditional invariant definition $(P(\bm{x}) \Longleftrightarrow u)$, which clears the ambiguity for $u$ value when $P(\bm{x})$ is false. This could be implemented in software using an IF THEN ELSE construct on the form: $\texttt{IF } P(\bm{x}) \texttt{ THEN } u=1 \quad \texttt{ELSE } u=0$.

A simple approach to design an attack out of safety requirements is to negate the invariant:
\begin{align}
    \lnot \left( P(\bm{x}) \Longleftrightarrow u \right) \equiv \left( P(\bm{x}) \Longleftrightarrow \lnot u \right)
    \label{eq:invariant-negated}
\end{align}
which has the software implementation:
\begin{align}
    \texttt{IF } P(\bm{x}) \texttt{ THEN } u=0 \quad \texttt{ELSE } u=1 \nonumber
\end{align}
This is intuitive, because it implements the exact opposite of the safety property, which should be capable of producing a safety hazard in the system. We note that the right hand side in  (\ref{eq:invariant-negated}) has the logical equivalence:
\begin{align}
    \left( P(\bm{x}) \Longleftrightarrow \lnot u \right) \equiv \underbrace{\left( \lnot P(\bm{x}) \land u \right)}_{\text{Disruption}} \lor \underbrace{\left( P(\bm{x}) \land \lnot u \right)}_{\text{Hazard}}
    \label{eq:invariant-negated-equivalent}
\end{align}
which could be interpreted as two ways to inject the attack. The first way, represented by the first term on the right hand side in (\ref{eq:invariant-negated-equivalent}), enforces the actuator value without the cause $P(\bm{x})$ being present. This results in a system disruption, but no hazard is involved since the safe action has already been taken by enforcing the safe actuator value. The second way, represented by the second term on the right hand side in (\ref{eq:invariant-negated-equivalent}), prevents the safe actuator action from being taken while the cause $P(\bm{x})$ being present. This results in a hazard as it violates the safety property. Therefore, the utilization of a bi-conditional invariant definition results in two possible attack scenarios.

\begin{exmp} \label{ex:pump-control-attacks}
Consider again the pump safety property in Example \ref{ex:pump-control}. Applying the attack design methodology by negating the invariant results in the software implementation:
\begin{align}
    & \texttt{IF } (x_1 < 10\%) \lor (x_2 > 5) \texttt{ THEN } u=1 \quad (\text{Hazard}) \nonumber \\
    & \texttt{ELSE } u=0 \quad (\text{Disruption}) \nonumber
\end{align}
This code results in two effects: \begin{inparaenum}[(1)] \item System hazard: resulting from unsafe pump operation. When the tank level is low, the pump will turn on, potentially resulting in pump cavitation damage. In addition, if the pump discharge pressure is high, possibly due to a blockage in the discharge pipe, the pump will turn on, causing a mounting pressure and potential rupture in the piping system. \item System service disruption: when both tank level and discharge pressure are normal, the code stops the pump, although the pump is supposed to operate to transfer the product in the tank.\end{inparaenum} \hfill $\blacksquare$
\end{exmp}

\subsection{Disruption Attack}
The prerequisite system state for the disruption attack, $\lnot P(\bm{x})$, is essentially the absence of a state condition that requires a safety action. In other words, it represents the normal state of the system. Therefore, this prerequisite system state will almost always be available for the attacker. The attacker does not need to enforce a specific system state in order to launch the attack, as in the case of hazard attacks discussed below. Rather, the attacker will just reverse the actuator output to cause system disruption. Although this seems as a soft target for the attacker, nothing comes for free. The consequence of the attack will just be a process disruption, and its associated financial loss. However, no safety hazard will be expected from such an attack. Moreover, it is hard to conceal a disruption attack for a long time from system operators, as the disruption of one system component propagates quickly to other connected components, and the attacker's task to conceal the attack would require faking many state variables simultaneously. This requires a global system knowledge, which is typically not available for the attacker. Regardless, the disruption attack is still a viable scenario for the penetration tester to assess CPS resilience against a highly-probable attack.

\subsection{Hazard Attack}
The hazard attack requires the specific system state condition as per the safety specification, $P(\bm{x})$. Therefore, the attacker should drive the system to the required state to launch the attack. This is harder than the disruption attack because it requires some system knowledge. Of course, the attacker could design the malware as dormant waiting for this specific system state to be reached to launch the attack. This specific system state may take years to be reached. Therefore, in the rest of the paper, we will discuss the proactive attack approach, where the attacker tries to drive the system to the desired state.

We classify system states into physical states, environmental states, and operator action states, i.e., $\bm{x} = [\bm{x}_p \quad \bm{x}_e \quad \bm{x}_o]$. The physical state $\bm{x}_p$ represents CPS measurements, such as car velocity, process tank level, or pump on/off status. The environmental state $\bm{x}_e$ typically represents disturbances affecting the CPS performance such as wind and temperature. The operator actions state $\bm{x}_o$ represents inputs from the operator to manually control the system. Among these three categories, the operator action is the easiest because the attacker can simply enforce the operator action variables to the desired values. The environmental states are out of control of both the system operator and the attacker. Therefore, the attacker can do nothing about it if the system state condition includes environmental states. For the physical state $\bm{x}_p$, if it is a system component state, it could be enforced by directly injecting an actuator attack. However, if the state represents system physical measurements that need to be enforced, then a sensor injection attack will not work, the physical system itself has to be driven to generate the desired state. This needs a more in-depth discussion.

In cyber physical systems, the system state could be changed by manipulating system inputs $\bm{u}$, also known as actuator values. Knowledge about system dynamics are needed to force the states to specific values. This knowledge is typically not available to the attacker. However, the relationship between system inputs $\bm{u}$ and states $\bm{x}_p$ could be inferred from the control algorithms implemented on the system controllers. In this paper, we give an example for Proportional Integral Derivative (PID) controllers that are pervasively used in cyber physical systems. However, the discussion could be equally adopted to other control algorithms.

The PID controller is a single-input single-output controller that works on the error between a setpoint and system state measurement. The error is used to calculate the proportional, integral, and derivative terms of the controller, which are fused together to calculate the controller output that will drive the system input actuator to bring the system state as close as possible to the required setpoint. Figure \ref{fig:Tank-Pump} illustrates an example PID controller feedback control system to regulate the tank level. The PID controller algorithm is the simple, yet powerful, equation:
\begin{align}
u(t) &= K_p e(t) + K_i \int_0^t{e(t) dt} + K_d \frac{de(t)}{dt}    \\
e(t) &= \text{SP}(t) - \text{PV}(t)
\end{align}
where SP is the setpoint, PV is the process variable, and $K_p, K_i, K_d$ are the controller parameters, which are tuned according to the controlled system. In order for the PID controller to work properly, it has to form a \emph{negative} feedback loop with the system. In other words, as the error increases, the controller output should select the right direction of change, either increase or decrease, in order to reduce the error. This is called in industry "Controller Action", and depends on how the system dynamics work and on the actuator. If the controller output increases as the error increases, it is called "Direct Acting" controller. On the other hand, a "Reverse Acting" controller decreases its output as the error increases. This is best illustrated by an example before explaining how to exploit this knowledge to launch an attack.

\begin{exmp}
Consider the Tank-Valve system in the left hand side of Figure \ref{fig:Tank-Pump}. As the tank level deviates from the setpoint, the valve opens or closes to bring the level back to its desired setpoint according to the PID controller algorithm. Suppose initially that when the PID control signal is 0\%, the valve is fully-closed, and when it is 100\%, the valve is fully-open (called fail-closed valve in industrial literature). Now, as the level increases above the setpoint, the error decreases, and we need to close the valve further to reduce the tank level, i.e., we need to decrease the PID control signal. Therefore, the controller has to decrease the output when the error increases and vice versa, i.e., it should be set as "Direct Acting".

Now suppose for safety reasons, the valve is designed such that when the PID controller output is 0\%, the valve is fully-open, and when the output is 100\%, the valve is fully-closed (called fail-open valve in industrial literature). With this system configuration, when the tank level increases, the error decreases, and we need to close the valve further, i.e., increase the PID control output. In such case, the PID controller should increase its output when the error decreases and vice-versa, i.e., it should be set as "Reverse Acting". \hfill $\blacksquare$
\end{exmp}
In practice, changing direct and reverse acting setting is done via swapping the setpoint and process variable terms in the error definition. Notice that in some industrial literature, the error is defined as $e = \text{PV} - \text{SP}$, and therefore, the logic of the above discussion will be reversed. However, the core idea remains the same.

The PID controller setting enables us to infer the system dynamics without knowing any information about the system. If the PID is Direct Acting, we know that as the error increases (equivalently the measurement decreases), the controller output increases. Therefore, if the attacker wants to increase this specific system measurement, the associated controller output needs to be decreased, and vice versa. This gives the attacker a systematic approach to drive the system states to the desired state condition expressed by $P(\bm{x})$ in order to launch a hazard attack. This knowledge could be obtained by searching through the control program.

\begin{exmp}
Consider an expanded version of the tank-pump system in Example \ref{ex:pump-control}, now including the inlet pipe and valve and a feedback PID control loop that regulates the inlet valve opening to control the tank level, as depicted in Figure \ref{fig:Tank-Pump} at the tank inlet. Assuming the valve opens with increasing controller signal, as the level decreases below the setpoint, the error increases, and the valve needs to open to restore the level, i.e., controller output needs to increase. Therefore, we need a direct acting PID controller. Therefore, by inspecting the PID configuration, the attacker knows that in order to decrease the level, the valve needs to be closed. Recalling that the safety property in Example \ref{ex:pump-control} includes $x_1 < 10\%$, then the attacker's objective is to reduce the level, hence fully closes the valve by setting the PID controller output to 0. \hfill $\blacksquare$
\end{exmp} 
Algorithm \ref{alg:malware} is a pseudo code that automates the malware payload development for a hazard attack, where safety properties are encoded using \texttt{IF...ELSE} constructs in structured text programming language, and assuming PID control algorithms. It takes as input the safety and control program files, searches for the invariants and control information, and returns malware payload source code files. This payload could then be encapsulated in a malware file that uses any of the common techniques to exploit a controller vulnerability to execute the payload, e.g., buffer overflow.
\begin{algorithm}[tb!]
\small
\SetAlgoLined
\SetKwInOut{Input}{Input}
\SetKwInOut{Output}{Output}
\Input{$F_s,F_c$ \qquad (Safety and Control code files)}
\Output{$F_{ms}, F_{mc}$ \qquad (Malware payload code files)}
$\bm{P(\bm{x}}), u, \text{IF-Body} \leftarrow$ Search($F_s$, "IF $\lor$ IF*ELSE") \\
Mal-Text1 $\leftarrow$ Replace(IF-Body, $*=1,*=0$) \\
Mal-Text2 $\leftarrow$ Replace(IF-Body, $*=0,*=1$) \\
Write($F_{ms}$, Mal-Text1, Mal-Text2) \\
\For{$P \in \bm{P}(\bm{x})$}{
\For{$x \in \bm{x}$}
{
    $C_x, u_x \leftarrow$ Search($F_c,x$) \quad (Controller ID, output) \\
    \eIf{Condition($x,P(\bm{x})$ == "$>$")}
    {
        \eIf{DA($C_x$)}
        { 
            Append($F_{mc}$, "$u_x = \min(u_x)$")
        }
        {
            Append($F_{mc}$, "$u_x = \max(u_x)$")
        }
    }
    {
        \eIf{RA($C_x$)}
        {
            Append($F_{mc}$, "$u_x = \max(u_x)$")
        }
        {
            Append($F_{mc}$, "$u_x = \min(u_x)$")
        }
    }
}
}
\Return $F_{ms},F_{mc}$
 \caption{Malware Payload Development for hazard attacks. Direct Acting DA(.) function returns True if the controller is direct acting, and similarly for Reverse Acting RA(.)}
 \label{alg:malware}
\end{algorithm}
\section{Attack Injection} \label{sec:AttackInjection}
\begin{figure}[tb!]
    \centering
    \includegraphics[scale=0.45]{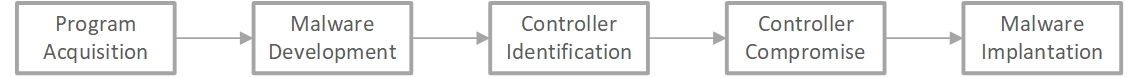}
    \caption{Attack injection process}
    \label{fig:Attack-Injection-Process}
\end{figure}

The attack injection process is illustrated in Figure \ref{fig:Attack-Injection-Process}. The process starts with searching and locating the safety and control programs, followed by the development of a malware as presented in this paper. The implementation controller is then located and compromised to gain access control, which typically involves privilege escalation. Finally, the running safety and control codes are disabled and replaced by the malicious codes, optionally launching the attack in stealthy mode by generating fake normal data to the human display and logging nodes. In the following, we briefly explain each attack step.

\subsection{Safety and Control Programs Acquisition}
One of the key questions is how to acquire the safety and control programs. After all, the technique presented in this paper relies mainly on the availability of these programs. It may be tempting to believe that a simple information security policy could prevent the leakage of these critical software programs. Although this is partially true, the reality is far less ideal. There are multiple stakeholders for these software programs. Examples are plant operators and engineers, automation vendor engineers, consultancy firms that originally designed the system, and the construction company. For operability reasons, stakeholders must keep the latest version of the software. Even worse, the turn around time for stakeholder engineers is so high that an engineer working today for company X may move to company Y few months later, with a copy of the program on his laptop or flash memory. Controlling the distribution of the program beyond the system owner's perimeter is not that straightforward, and a trade-off has to be made between security and operation convenience as well as fast incident resolution.

One of the typical problems in embedded systems programming is maintaining the source code of the deployed program on the controller. To solve this issue, some of the vendors introduced a source code program storage capability on the controller itself, i.e., when the code is compiled and downloaded to the controller, the source code is downloaded as well and permanently stored on the controller. This gives an additional opportunity for the attacker to obtain the source code once the controller is compromised. 

\subsection{Malware Development}
The development of malware for a given CPS is described in Section \ref{sec:AttackDesign} and summarized in Algorithm \ref{alg:malware}.

\subsection{Locating the Controller}
After obtaining the software, it is required to identify which controller executes the code. Different software tools have different methods to reference the controller in the source code. Typically, the controller IP address or name tag will be referenced in the source code.

\subsection{Controller Compromise}
This is a pre-requisite attack step before malware implantation. The controller has to be compromised with escalated privileges. The compromise of a CPS controller in an operating environment is an active area of research and depends mainly on the CPS domain and cyber system architecture. The interested reader may refer to an example in \cite{Tantawy2020} for a process control plant compromise or consult the literature for the specific CPS domain of interest.

\subsection{Malware Implantation}
\begin{figure}
    \centering
    \includegraphics[scale=0.75]{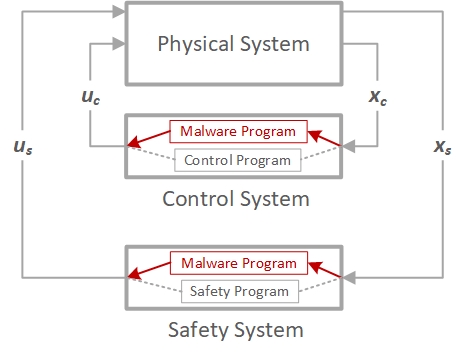}
    \caption{Malware injection into a CPS. Simultaneous injection into the control and safety systems may lead to a hazard attack, while safety system injection may result in a disruption attack.}
    \label{fig:Malware-Injection}
\end{figure}

This process is illustrated in Figure \ref{fig:Malware-Injection}. The existing code has to be disabled, and the malware has to be loaded in the controller memory. An alternative approach is to launch the malware only when the desired system state is reached, i.e., the dormant attack discussed in the paper. In both cases, the attacker may attempt to generate fake data to the system observing nodes to cover the attack, although this is more challenging.

\section{Discussion}
The paper proposes the idea that a CPS attacker does not need to understand the system dynamics or behavior in order to launch a destructive attack. This is because the required information has already been extracted by system designers and encoded, partially or fully, in the control and safety algorithms. The likelihood of an attack will depend on the availability of the required data in the control code as well as the difficulty of decoding such information into system knowledge that could be used by the attacker. The method presented in Algorithm \ref{alg:malware} is for structured text programs that encode the safety properties using \texttt{IF...ELSE} constructs. Although quite common in industry, this is not the only method to encode and program safety properties. A more general approach is clearly needed to extract safety-critical attacks from the diverse representations and programming languages.

One important question is how we can utilize the fact that attackers can reverse-engineer our system to better secure it? There are several ways: \begin{inparaenum}[(1)] \item we can carry out a more informed risk assessment process, focusing on the most relevant attacks, and design the most suitable security mechanisms. This is enabled by the fact that we know what attacks are harmful, how much damage they could cause, and their likelihood that is proportional to the difficulty of decoding the relevant information, \item better design of penetration testing by including more relevant safety-critical attacks to measure the system robustness and identify any system vulnerabilities, \item the information about safety-critical attacks could be used during design time to develop more-resilient system architectures. As an example, if we know that an equipment is critical for the overall system safety, we can try to separate the hardware and software components that monitor and control such equipment from the rest of the control and safety systems, and put more security protection mechanisms for the isolated hardware and software. Even without an isolated hardware, we can still isolate the specific software task scheduled by the controller operating system by applying more strict access control rights. Both techniques are obviously more cost-effective and less human-demanding than trying to invest in securing the whole system. Such techniques could also be considered examples of the \emph{secure by design} principle.\end{inparaenum}

The proposed approach results in extreme attack consequences, i.e., a service disruption or system hazard. These consequences will be eventually noticed, either by system operators or by automated response systems. The impact of such an attack will vary significantly depending on other non-cyber protection and mitigation measures. On the other hand, incipient attacks intend to gradually damage the system, and therefore are mostly launched in a stealthy mode. Incipient attacks are harder to design, because they require more knowledge about how the system works. The extraction of such information from available control and safety programs is currently under investigation.

\section{Conclusion} \label{sec:Conclusion}
We presented a systematic method to automatically develop a CPS malware payload out of safety and control programs, without knowledge about the CPS. The two types of attacks presented are disruption attacks and hazard attacks. Disruption attacks are easier to implement, hence have more likelihood, but have less consequence severity. Hazard attacks are harder to implement, therefore are less likely, but have higher consequence severity. Several extensions could be identified out of this work. A more general algorithm that automates malware payload development for diverse representations is needed. Experimental evaluation of the approach on real control code and controllers of a CPS testbed is important to estimate the difficulty in reverse engineering and malware injection. In contrast with immediate effect attacks presented, automated design of incipient attacks by reverse engineering is important during both design and operation times. Finally, learning system dynamics from sensor and actuator communication by network eavesdropping in order to craft an attack is important when control programs cannot be obtained.



\bibliographystyle{ieeetr}

\bibliography{references2.bib}

\end{document}